\documentclass[epj]{svjour}
\usepackage{latexsym}
\usepackage{array}
\usepackage{amsmath}
\usepackage{epsfig}
\usepackage{graphics}
\usepackage{booktabs}
\usepackage{feynmp}
\def\lsim{\mathrel{\rlap{\lower5pt\hbox{\hskip1pt$\sim$}}
\raise5pt\hbox{$<$}}}
\def\gsim{\mathrel{\rlap{\lower5pt\hbox{\hskip1pt$\sim$}}
\raise5pt\hbox{$>$}}}
\newcommand{\mrm}[1]{\mathrm{#1}}
\newcommand{\pT}{p_{\perp}}                  
\newcommand{\kT}{k_{\perp}}

\renewcommand{\c}{\mathrm{c}}
\renewcommand{\d}{\mathrm{d}}

\newcommand{\g}{\mathrm{g}}

\newcommand{\q}{\mathrm{q}}
\newcommand{\s}{\mathrm{s}}

\newcommand{\qbar}{\overline{\mathrm{q}}}

\newcommand{\qsea}{\ensuremath{\q_{\mrm{s}}}}
\newcommand{\qcmp}{\ensuremath{\q_{\mrm{c}}}}
\newcommand{\val}{\ensuremath{{\mrm{v}}}}
\newcommand{\sea}{\ensuremath{{\mrm{s}}}}
\newcommand{\cmp}{\ensuremath{{\mrm{c}}}}
\begin{document}
\title{Progress on Multiple Interactions}
\subtitle{Modelling the underlying event in hadron--hadron collisions}
\author{P.~Skands (speaker) and T.~Sj\"ostrand% etc
% \thanks is optional - remove next line if not needed
%\thanks{\emph{Present address:} Insert the address here if needed}%
}                     % Do not remove
%
%\offprints{}          % Insert a name or remove this line
%
\institute{Theoretical Physics, Lund University, S\"olvegatan 14A, 22362
  Lund, Sweden.}
\date{LU TP 03-45, hep-ph/0310315}
\abstract{We report on the development of a new model for the underlying
  event in hadron--hadron collisions. The model
  includes parton showers for all interactions, as well as
  non-trivial flavour, momentum, and colour correlations between 
  interaction initiators and beam remnant partons.\\
 % end of PACS codes
} %end of abstract
\maketitle
\section{Introduction} 
A simple consequence of the composite nature of hadrons is the
possibility to have hadron--hadron collisions in which several distinct pairs of
partons collide with each other (multiple interactions). In fact, simple
perturbative calculations can be used to show \cite{Sjostrand:1987su} that
most inelastic events in high--energy hadronic collisions should
contain several perturbatively calculable interactions, in addition to whatever
nonperturbative phenomena may be present.   

Although most of this activity is not hard enough to play a
significant role in the description of high--$\pT$ jet physics, it can be
responsible for a large fraction of the total multiplicity (and large
\emph{fluctuations} in it), for semi-hard (mini-)jets in the   
event, for the details of jet profiles and for the jet pedestal effect,
leading to random as well as systematic shifts in the jet energy scale. Thus, a
good understanding of multiple interactions would seem prerequisite 
to carrying out precision studies involving jets and/or the underlying event
in hadronic collisions. 

In an earlier study \cite{Sjostrand:1987su}, it was
argued that \emph{all} the underlying event activity could be explained by the
multiple interactions mechanism alone. However, while the origin of
underlying events is thus assumed to be perturbative, many nonperturbative
aspects still force their entrance on the stage. This in particular relates
to the structure of beam remnants and to the correlations in flavour, colour, and
momentum between the partons involved. In \cite{Sjostrand:1987su}, 
only very simple beam remnant structures could technically be dealt with,
hence substantial simplifications had to be imposed.

In recent years, the physics of the underlying event has come to attract more
attention. Simple parameterizations can be tuned to describe the average
underlying activity, but are inadequate to fully describe correlations and
fluctuations. The increased interest and the new data now prompts us to
develop a more realistic framework for multiple interactions than the one in
ref.~\cite{Sjostrand:1987su}, while making use of many of the same underlying
ideas.

One new aspect was the augmentation in \cite{Sjostrand:2002ip} of the standard
Lund string fragmentation framework \cite{Andersson:1983ia} to include 
the hadronization of colour topologies containing non-zero baryon
number. In the context of multiple interactions, this improvement means that
almost arbitrarily complicated baryon beam remnants may now be dealt with,
hence many of the restrictions present in the old model are no longer necessary.
 
Here, we present a model
for how flavours, colours, and momenta are correlated between all
the partons involved in a hadron--hadron collision, 
both those that undergo interactions and those
constituting the beam remnants. However, all aspects of the model cannot be
treated within the limits of this format, hence 
some aspects have been left out;  we concentrate exclusively
on baryon beams and neither the dependence on impact parameter nor the
assignment of `primordial $\kT$' to parton shower initiators is
addressed here. More 
complete descriptions may be found in \cite{Sjostrand:2003wg,inprep}.

%Clearly, the model we present here is not the final word. There are
%many unsolved issues left. For instance, we defer for the future 
%any discussions whether and how diffractive topologies could arise 
%naturally from several interactions with a net colour singlet exchange.
%The model also allows some options in a few places. A reasonable range 
%of possibilities can then be explored, and (eventually) experimental
%tests should teach us more about the course taken by nature.

This article is organized as follows. In Section~2 the main work on flavour
and momentum space correlations is presented and in 3 the very thorny issue 
of colour correlations. Finally, Section~4 provides 
a brief summary.
                   
\section{Towards a Realistic Model}
Consider a hadron undergoing multiple interactions in a collision. 
Such an object should be described by multi-parton densities, 
giving the joint probability of simultaneously finding $n$ partons with 
flavours $f_1,\ldots,f_n$, carrying momentum fractions $x_1,\ldots,x_n$ 
inside the hadron, when probed by interactions at scales 
$Q_1^2,\ldots,Q_n^2$. However, we are nowhere near having sufficient 
experimental information to pin down such distributions. Therefore, and
wishing to make maximal use of the information that we \emph{do} have, namely
the standard one-parton-inclusive parton densities, we propose the following
strategy. 

As described in \cite{Sjostrand:1987su}, the interactions may be generated in
an ordered sequence of falling $\pT$. For the hardest interaction, all
smaller $\pT$ scales may be effectively integrated out of the (unknown) fully
correlated  distributions, leaving an object described by the standard
one-parton distributions, by definition. For the second and subsequent
interactions, again all lower--$\pT$ scales can be integrated out, but the
correlations with the first cannot, and so on. 

Thus, we introduce modified parton densities, that correlate the $i$'th
interaction and its shower evolution to what happened in the $i-1$ previous
ones. 

The first and most trivial observation is that each interaction $i$
removes a momentum fraction $x_i$ from the hadron remnant. Already in
\cite{Sjostrand:1987su} this momentum loss was taken into account by assuming
a simple scaling ansatz for the parton distributions, $f(x) \to f(x/X)/X$,
where $X = 1 - \sum_{i=1}^n x_i$ is the momentum remaining in the beam hadron
after the $n$ first interactions. Effectively, the PDF's are simply `squeezed' into the range
$x\in[0,X]$. 

Next, for a given baryon, the valence
distribution of flavour $f$ after $n$ interactions, $q_{f\val n}(x,Q^2)$, 
should follow the counting rule:
\begin{equation}
  \int_{0}^{X}q_{f\val n}(x,Q^2)~\d x = N_{f\val n}, \label{eq:valnumrule}
\end{equation}
where $N_{f\val n}$ is the number of valence quarks of flavour $f$ remaining
in the hadron remnant. This rule may be enforced by scaling the original
distribution down, by the ratio of remaining to original valence quarks
$N_{f\val n}/N_{f\val 0}$, in addition to the $x$ scaling mentioned above.

Also, when a sea quark is knocked out of a hadron, it must leave behind a
corresponding antisea parton in the beam remnant. We
call this a companion quark. 
In the perturbative approximation the sea quark $\qsea$ and its
companion $\qcmp$ come from a gluon branching
$\g \to \qsea + \qcmp$ (it is implicit that if \qsea\ is a quark, \qcmp\ is
its antiquark). Starting from this perturbative ansatz, and 
neglecting other interactions and 
any subsequent perturbative evolution of the $q_{\cmp}$ 
distribution, we obtain the $q_{\cmp}$ distribution from
the probability that a sea quark $\qsea$, carrying a
momentum fraction $x_{\sea}$, is produced by the branching of a
gluon with momentum fraction $y$, so that the 
companion has a momentum fraction $x=y-x_{\sea}$, 
\begin{eqnarray}
q_{\cmp}(x;x_{\sea}) & = & C\int_0^1
g(y) \, P_{\g\to\qsea\qcmp}(z) \, \delta(x_{\s}-zy)~\d z 
\nonumber \\
& = & C~g(y) \, P_{\g\to\qsea\qcmp}\left(\frac{x_{\s}}{y}\right)\frac{1}{y}
\nonumber \\
& = & C~\frac{g(x_{\s}+x)}{x_{\s}+x} \, 
P_{\g\to\qsea\qcmp}\left(\frac{x_{\s}}{x_{\s}+x}\right), 
\label{eq:qcmp}
\end{eqnarray}
with $P_{\g\to\qsea\qcmp}$ the usual DGLAP gluon splitting kernel and 
$C$ a normalization constant which can be obtained by imposing the counting 
rule:
\begin{equation}
\int_0^{1-x_{\sea}}\!q_{\cmp}(x;x_{\sea})~\d x = 1.
\end{equation}
The exact form of $C$ depends on the shape assumed for the gluon
disitribution. Qualitatively, however, any falling gluon
distribution $\propto 1/x$ convoluted with the almost flat $\g\to\q\qbar$
splitting kernel yields companion distributions which tend to a constant
$q_{\cmp}(x;x_s)\sim C/2x_{\s}^2$ below $x_{\s}$ and which exhibit power-like
fall-offs $q_{\cmp}(x;x_s)\propto 1/x^2$ above it, with some modulation of
the latter depending on the large-$x$ behaviour assumed for $g(x)$. Also note
that $xq_{\cmp}(x;x_s)$ should be peaked around $x \approx x_{\sea}$, by
virtue of the symmetric $P_{\g\to\qsea\qcmp}$ splitting kernel.

Without any further change, the reduction of the valence
distributions and the introduction of companion distributions, in the manner
described above, would result in a violation of the total momentum sum rule, 
\begin{equation}
  \int_0^{X}\!x\left(\sum_f q_{fn}(x,Q^2) + g_n(x,Q^2)\right)\d x =
  X, \label{eq:totmomrule} 
\end{equation}
since by removing a valence quark from the parton distributions 
we also remove a total amount of momentum corresponding to 
$\langle x_{f\val} \rangle$, the average momentum fraction carried by a 
valence quark of flavour $f$:
\begin{equation}
\langle x_{f\val n} \rangle \equiv \frac{\int_0^X xq_{f\val n}(x,Q^2)~\d x}%
{\int_0^X q_{f\val n}(x,Q^2)~\d x} = X \, \langle x_{f\val 0} \rangle ~,
\label{eq:avmom}
\end{equation}
and by adding a companion distribution we add an analogously defined
momentum fraction. 

To ensure that eq.~(\ref{eq:totmomrule}) is still respected, we assume that 
the sea+gluon normalizations fluctuate up 
when a valence distribution is reduced and down when a companion
distribution is added. In addition, the requirement of a physical $x$ range
is of course still maintained by `squeezing' all distributions into the interval
$x\in[0,X]$. 

The full parton distributions after $n$ interactions thus take the forms:
\begin{eqnarray}
{q_{f n}\left(x\right)} & = & 
 \frac{1}{X}\left[ \frac{N_{f\val n}}{N_{f\val 0}}
{q_{f\val 0}\left(\frac{x}{X}\right)}\right.+ a\,
q_{f \sea 0} \left(\frac{x}{X}\right) + \nonumber \\ & & + \left.
\sum_{j} q_{f\cmp_j}
\left(\frac{x}{X};x_{s_j}\right) \right]  \, ,
\\
\displaystyle {g_n(x)} &=&\frac{a}{X} g_0\left(\frac{x}{X}\right),
\end{eqnarray}
where we have suppressed the dependence on $Q^2$ for brevity, 
$q_{f\val 0}$ ($q_{f \sea
  0}$) denotes the original valence (sea) distribution of flavour $f$, and the
  index $j$ on the companion distributions $q_{f\cmp_j}$ counts different
  companion quarks of the same flavour, $f$. As already mentioned, 
the normalization factor $a$ multiplying the gluon and sea distributions can be 
determined from overall momentum conservation in the incoming
hadron:
\begin{equation}
a = \frac{1-\sum_fN_{f\val n}\langle x_{f\val 0} \rangle
-\sum_{f,j} \langle x_{f\cmp_j 0} \rangle}{1- \sum_fN_{f\val 0}\langle x_{f\val
    0} \rangle}. 
\end{equation}

After the perturbative interactions have taken each their share of
longitudinal momentum, the question arises how the remaining momentum
is shared between the beam remnant partons. 
Here, valence quarks receive an $x$ picked at random according to a 
small--$Q^2$ valence-like parton density, while sea quarks must be 
companions of one of the initiator quarks, and hence should 
have an $x$ picked according to the $q_{\c}(x ; x_{\s})$ distribution 
introduced above. In the rare case that no valence quarks remain and no sea 
quarks need be added for flavour conservation, the beam remnant is 
represented by a gluon, carrying all of the beam remnant longitudinal 
momentum. 

Further aspects of the model include the possible formation of composite
objects in the beam remnants (e.g.~diquarks) and the addition
of non-zero primordial $\kT$ values to the parton shower
initiators. Especially the latter introduces some complications, to
obtain consistent kinematics. Details on these aspects 
will be presented in \cite{inprep}. 

\section{Colour Correlations}
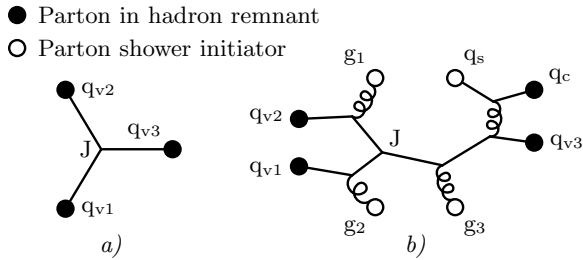
\begin{figure}
\begin{center}\vspace*{3mm}
\begin{fmffile}{fmfinitial}
\begin{tabular}{ccc}
\raisebox{1.7cm}{\hspace*{-1.3cm}\begin{fmfgraph*}(50,25)
\fmftop{t}
\fmfbottom{b}
\fmf{phantom}{t,v,b}
\fmfv{d.sh=circ,d.f=full,d.siz=6,lab=Parton in hadron
  remnant,l.ang=0,l.dist=8}{t} 
\fmfv{d.sh=circ,d.f=empty,d.siz=6,lab=Parton shower initiator,l.ang=0,l.dist=8}{v} 
\end{fmfgraph*}}\hspace*{-0.5cm}
%\begin{fmfgraph*}(60,60)
%\fmfforce{0.5w,0.5h}{j}
%\fmfv{d.sh=circ,d.f=20,d.siz=1.4h}{j}
%\end{fmfgraph*}\hspace*{-58.4\unitlength}
\begin{fmfgraph*}(45,45)
\fmfleft{q1,q2}
\fmfright{q3}
\fmf{plain}{q1,j}
\fmf{plain}{q2,j}
\fmf{plain}{q3,j}
\fmfv{d.sh=circ,d.f=full,d.siz=6,lab=$\q_{\val1}$,l.ang=0}{q1}
\fmfv{d.sh=circ,d.f=full,d.siz=6,lab=$\q_{\val2}$,l.ang=0}{q2}
\fmfv{d.sh=circ,d.f=full,d.siz=6,lab=$\q_{\val3}$,l.ang=135}{q3}
\fmfv{lab=J,l.ang=180,l.dist=3.5}{j}
\end{fmfgraph*} &\hspace*{0.8cm} &
\begin{fmfgraph*}(90,50)
\fmfset{curly_len}{2.2mm}
\fmftop{d1,g1,t2,t3}
\fmfbottom{d3,g2,g3,d4}
\fmfleft{d5,q1,q2,d6}
\fmfright{q3}
\fmf{plain}{q1,vq1,j}
\fmf{plain}{q2,vq2,j}
\fmf{plain}{q3,vq3,vq4,j}
\fmf{gluon}{vq1,g2}
\fmf{gluon}{vq2,g1}
\fmf{gluon}{vq3,vqq}
\fmf{plain}{t2,vqq,t3}
\fmf{gluon}{vq4,g3}
\fmfv{d.sh=circ,d.f=full,d.siz=6,lab=$\q_{\val1}$}{q1}
\fmfv{d.sh=circ,d.f=full,d.siz=6,lab=$\q_{\val2}$}{q2}
\fmfv{d.sh=circ,d.f=full,d.siz=6,lab=$\q_{\val3}$}{q3}
\fmfv{lab=J,l.ang=45,l.dist=3.5}{j}
\fmfv{d.sh=circ,d.f=empty,d.siz=6,lab=$\g_1$}{g1}
\fmfv{d.sh=circ,d.f=empty,d.siz=6,lab=$\g_2$}{g2}
\fmfv{d.sh=circ,d.f=empty,d.siz=6,lab=$\g_3$}{g3}
\fmfv{d.sh=circ,d.f=empty,d.siz=6,lab=$\qsea$}{t2}
\fmfv{d.sh=circ,d.f=full,d.siz=6,lab=$\qcmp$}{t3}
\end{fmfgraph*}
\\[2mm]
\it a) & & \it b)
\end{tabular}
\end{fmffile}
\caption{{\it a)} The initial state of a baryon, consisting of 3 valence quarks
  conneted antisymmetrically in colour via a central `string junction', J. 
{\it b)} Example of how a given set of parton shower intitators could have
  been radiated off the initial
configuration, in the case of the `purely random' correlations discussed in
  the text. 
\label{fig:initialstate}}
\end{center}
\end{figure}

The initial state of a baryon may be represented by three valence quarks,
connected antisymmetrically in colour via a central junction, which acts as a
switchyard for the colour flow and carries the net baryon number. 
This situation is illustrated in Fig.~\ref{fig:initialstate}a. 

The precise colour-space 
evolution of this state into the initiator and beam remnant partons
actually found in a given event is not predicted by perturbation theory, but is
crucial in determining how the system hadronizes; in the Lund string model
\cite{Andersson:1983ia},
two colour-connected final state partons together define a string piece,
which hadronizes by successive non-perturbative
breakups along the string. Thus, the colour flow of an event determines
the topology of the hadronizing strings, and consequentially
where and how many hadrons will be produced. 

For the perturbative parts of the event, a unique colour flow may be
consistently assigned \cite{Bengtsson:1984jr}, 
but for the connections among initiator and
beam remnant partons, additional assumptions are necessary. The question can
essentially be reduced to one of choosing a fictitious sequence of gluon
emissions off the initial valence topology, since sea quarks together with
their companion partners are associated with parent gluons, by definition.

The simplest solution is to assume that gluons are attached to the initial
quark lines in a random order, see Fig.~\ref{fig:initialstate}b. 
If so, the junction would rarely be colour connected directly to two valence
quarks in the beam remnant. It should be clear that the migration of the baryon
number depends sensitively upon which partons in the final state the junction
ends up being connected to (see \cite{Sjostrand:2002ip} for details on 
junction fragmentation). Thus, if the connections are \emph{purely} random, 
the baryon number of the initial state should quite often be
disconnected from the beam remnant altogether and be able to migrate to
both large $\pT$ and small $x_F$ values.  
Empirically, this may not be desireable, hence a free suppression parameter
is introduced to suppress gluon attachments onto colour lines that lie
entirely within the remnant. 

Finally, we imagine a few different possibilities for the ordering of the
emissions off a given colour 
line: 1) random, 2) gluons are ordered according to the
rapidity of the hard scattering subsystem they are associated with (where beam
remnant partons are assigned a fixed large rapidity in the direction of their
parent hadrons), and 3) gluons are ordered so as to give rise to the smallest
possible total string lengths in the final state. The two latter
possibilities represent different attempts to minimize the total potential
energy of the system (the string length), since the former seems to result in
a too large hadron multiplicity per interaction. A variable which we have 
found to be sensitive to the choice of colour topology is the average $\pT$
vs.~$n_{\mathrm{charged}}$, but our studies are not concluded yet. Thus,
we do not pretend to have the final solution to these questions. Rather, 
the colour correlations, both in the initial and in the final
state, still represents the major open issue in our studies. 

\section{Conclusion}
The development of a new model for the underlying event in hadron--hadron
 collisions  has been reported. This model extends the multiple 
 interactions mechanism proposed in \cite{Sjostrand:1987su} with the
 possibility of non-trivial flavour and momentum correlations, parton
 showers for all initiator and final state partons, and several options for
 colour correlations between initiator and beam remnant partons. Many of
 these improvements rely on the development of junction fragmentation in
 \cite{Sjostrand:2002ip}. 

 The issue of colour correlations is still actively under study. 

\bibliography{proc_skands}

\begin{thebibliography}{1}

\bibitem{Sjostrand:1987su}
T.~Sj{\"o}strand, M.~van Zijl,
\newblock Phys. Rev. {\bf D36}, 2019 (1987).
%%CITATION = PHRVA,D36,2019;%%

\bibitem{Sjostrand:2002ip}
T.~Sj{\"o}strand, P.~Z. Skands,
\newblock Nucl. Phys. {\bf B659}, 243 (2003).
%%CITATION = HEP-PH 0212264;%%

\bibitem{Andersson:1983ia}
B.~Andersson{ et~al.},
\newblock Phys. Rept. {\bf 97}, 31 (1983).
%%CITATION = PRPLC,97,31;%%

\bibitem{Sjostrand:2003wg}
T.~Sj{\"o}strand{ et~al.},
\newblock LU TP 03-38. [hep-ph/0308153]
%%CITATION = HEP-PH 0308153;%%

\bibitem{inprep}
T.~Sj{\"o}strand, P.~Z. Skands,
\newblock in preparation.

\bibitem{Bengtsson:1984jr}
H.~U. Bengtsson,
\newblock Comput. Phys. Commun. {\bf 31}, 323 (1984).
%%CITATION = CPHCB,31,323;%%

\end{thebibliography}
\end{document}